\documentstyle[twoside,fleqn,gracos]{article}
\jheads{V.V. Kassandrov}
       {Biquaternion Electrodynamics and Weyl-Cartan Geometry of Space-Time}

\def\B{\cal}
\def\Bb{\cal}
\def\Bbb{\cal}

\begin{document}
\twocolumn[
\Arthead{3}{216}{222}

\Title{BIQUATERNION ELECTRODYNAMICS\\[5pt]
	  AND WEYL-CARTAN GEOMETRY OF SPACE-TIME}

\Author{V.V.Kassandrov}
{Russian People's  Friendship  University, Department of General 
Physics,\\ 3 Ordjonikidze Str., Moscow 117302, Russia}

\Rec{5 July 1995}

\Abstract
{The generalized  Cauchy-Riemann  equations  (GCRE)  in  biquaternion
algebra appear to be Lorentz-invariant. The Laplace equation is in this case
replaced by a nonlinear {\B C}\,-eikonal equation. GCRE contain a 2-spinor
and a {\B C}\,-gauge structures, and their integrability conditions take the
form of Maxwell and Yang-Mills equations. For the value  of  electric
charge from GCRE only the quantization rule follows, as well  as  the
treatment of Coulomb law  as  a  stereographic  map.  The  equivalent
geometrodynamics in a Weyl-Cartan affine space and  the  conjecture of
a complex-quaternion structure of space-time are discussed.}

] 

\section{Introduction}               
In the frames of the {\it geometrodynamic\/}  approach  all
fundamental physical quantities and above all the  equations  of
physical dynamics should  be  of a purely  geometric  nature.  The twistor
program, the Ka\-lu\-za-Kle\-in  theories  and  string  dynamics  give
representative examples of this concept, perhaps the most general ones up
to now. In essence, any physical interaction may  be  regarded  as  a
manifestation of geometry (by using multi-dimensional spaces, fiber bundles,
etc.).

However,  the  diversity  of  admissible  geometries  and their
invariants makes the ``kinematic'' part of this procedure  (selection
of space and geometric identification of  physical  quantities),  as
well as the dynamic one (choice  of a Lagrangian)  quite  ambiguous.
Even for the electromagnetic (EM) field  one  has  a  lot  of  different
geometric interpretations  (Weyl's  conformal  factor,  bundle
connection, the Kaluza metric field,  torsion  [1]  or  nonholonomic [2]
structures of space-time (ST), etc.).

Alternatively, within the {\it algebrodynamic\/} pa\-ra\-di\-gm [3] the ST is
regarded as a manifold supplied with a basic algebraic structure, the
structure of linear algebra in the simplest case. But it is well known
that the exceptional algebras --- algebras with  division  and  positive
norm --- exist in the dimensions $d=4$ (Hamilton quaternions) and $d=8$
(Cayley octonions). So it  would  be  natural  to suppose  that  the
ST algebra  (STA)  [4]  should  be  exceptional  in  its  internal
mathematical properties. If it is the case, the group of automorphisms
(Aut) of STA would generate the ST geometry,  for  example, by
operating as an isometry group.

Moreover,  the  STA structure  can  completely  determine
physical dynamics as well. Indeed, if we consider the physical fields
as algebra-valued functions of an algebraic  variable,  the {\it generalized
Cauchy-Ri\-emann equations\/} (GCRE), i.e., the differentiability conditions
in the STA, become fundamental  equations  of  field  dynamics.
Wonderfully, the generally accepted physical equations (in
particular, the Maxwell or Yang-Mills equations) become a direct
consequences  of GCRE, namely their integrability conditions (see below).

From an epistemological point of view, the  algebrodynamic  (AD)
concept returns us  to  the  ideas  of  Pythagoras,  Hamilton and
Eddington on {\bf a crucial role  of  Numbers  in  the  structure  of the
Universe}. At the modern stage we deal with the primary  structure  of
multidimensional {\it ST arithmetics},  completely  different  from  the
classical arithmetics of the Macroworld, or the world of  reversible
processes and weakly interacting objects.

{\bf A genuine ST arithmetics ought to be non-commutative and even non-%
associative!} Indeed, these properties are just algebraic  equivalents
of causal and interactive structures of the physical World (ensuring  the
dependence of ``out-state'' on the order and composition of reactions).
For such reasons the most  suitable  STA candidate  is  the {\it octonion
algebra}, the unique exceptional non-associative algebra. However, the
difficulties of ``intercourse''  with  octonions  are  well-known  (see,
nevertheless, [6\,7]).

Meanwhile, the non-commutativity of algebraic structures
is closely connected with the non-linearity  of the corresponding dynamic
equations (this is the case, in particular, for the Yang-Mills fields).
We will see later that the GCRE in non-commutative algebras also
possess  a nonlinear structure and are therefore capable  to describe both
quantum phenomena and physical field interactions.

In this paper we choose for a STA the {\it algebra of biquaternions\/}
{\Bbb B}\,,
the extension of real Hamilton quaternions {\Bbb H} to the field of complex
numbers {\Bbb C}. The {\Bbb H}\, algebra is known to have
${\rm Aut}\,(\mbox{\Bbb H})={\rm SO}\,(3)$ and  is  in
perfect correspondence with the structure of the 3-dimensional space.
{\bf We are unaware of a similar algebra for the  case  of  Minkowski
4-space!} For obvious reasons one often considers the Clifford-Dirac
algebra C(1,\,3) to be the STA [4,\,5].  However,  a reduction  from
the 16-dimensional total vector space of C(1,\,3) to a 4-dimensional
physical ST is  a completely ``voluntaristic'' procedure; even the metric
signature of the basic generator space may be chosen in different ways [8].

The {\Bbb B}-algebra, isomorphic to the Clifford algebra~C(3,0)  of
smaller dimension $d=8$, is preferable from this point of view.  On  the
other hand, the {\Bbb B}\,-dynamics, based on GCRE, appears to be Lorentz
invariant, so {\bf the {\Bbb B}\,-algebra may be treated as a minimal STA}.
This choice leads  to the conjectures on a fundamental role of {\it null
divisors\/} as a subspace of STA and on complex-valued structure of ST;
these questions will be discussed below.

Now we are ready to present the contents of the paper.
In Sec. 2  we  begin with the  basic  definitions  of the {\Bbb B}\,-algebra
and {\Bbb B}\,-differentiability. The general problems of (bi-)quaternionic
analysis are also briefly  discussed.
Then, in Sec. 3,  after  preliminary
physical identifications, we demonstrate the 2-spinor structure of the
basic GCRE and obtain a complexified eikonal equation for each
component of the {\Bbb B}\,-field. Global symmetries of the model are
studied as well.

Sections 4 and 5 are devoted to {\Bbb B}\,-electrodynamics as the basic case
of {\Bbb B}\,-differentiability. Firstly (Sec. 4) the self-duality
conditions are obtained from  GCRE,  whence follow the  Maxwell  equations.
Gauge invariance of a model of special type is demonstrated in Sec. 5. From
the eikonal equation, a geometrical origin of the Coulomb law as a
stereographic projection becomes  evident,  and we get for the admissible
values of an electric charge $q =\pm 1$,  i.e., a quantization rule!

In Sec. 6 we  demonstrate  the  equivalence  of the theory  to
geometrodynamics in a complexified Weyl-Cartan space. A reduction  to
Minkowski  space  identifies the  magnetic  monopole  field  as  that of
torsion and the Coloumb electric one as the ST Weyl nonmetricity.
We conclude in Sec. 7 by the establishment  of  complex-valued  Yang-%
Mills equations as the  integrability  conditions  of  GRCE  and a
discussion of general consequences of a complex-quaternionic  structure
of physical space. Finally, we  discuss  the  relation  of the
AD approach to binary geometrophysics.

\section{{\Bb B}\,-algebra and {\Bb B}\,-differentiability}   

Let {\bf z}$\in${\Bbb M}\,(4,\,{\Bbb C}), ${\bf z}=\{z^\mu,\ \mu=0,1,2,3\}$
be an element of the complex vector
space {\Bbb M}\,(4,\,{\Bbb C}) of dimension  $d=4$. The function
\beq                                                
{\bf F}({\bf z})=\left\{F^\mu(z)\right\}=\left\{F^\mu(z^0,z^1,z^2,z^3)\right\}
\eeq
{\bf F}$\in${\Bbb M}, maps an open domain {\Bbb O}$\subset${\Bbb M} to the
domain {\Bbb O}$\!{}^\prime\!\subset${\Bbb M}\,; let its components
$F^\mu(z)$ be complex and analytic.

Then a structure {\Bbb B} of associative algebra of complex quaternions
(biquaternions) {\Bbb M}$\times${\Bbb M}$\to${\Bbb M} may  be introduced  on
{\Bbb M}\,.  According  to the
isomorphism {\Bbb B}\ =\ {\Bbb L}(2,\,{\Bbb C}), {\Bbb L} being the full
2$\times$2 complex matrix algebra, we
shall use the matrix representation of {\Bbb B}
\beq                                
\forall{\bf z}\in\mbox{{\Bbb M}} : {\bf z}=z^\mu\sigma_\mu=
\left\|\begin{array}{cc}
u&w\\p&v\end{array}\right\|,
\eeq
$\sigma_\mu=\{e,\sigma_a\}$, $e$ being the unit 2$\times$2 matrix and
$\{\sigma_a,\ a=1,2,3\}$ the Pauli  matrices;
$u,v=z^0\pm z^3$; $p,w=z^1\pm iz^2$ are the DeWitt coordinates
on {\Bbb M}\,. Now the
multiplication $(\ast)$ in {\Bbb B} is equivalent to the  usual  matrix  one;
the function (1) becomes a matrix-valued, or {\Bbb B}\,-valued function of a
{\Bbb B}\,-variable. Let for some {\bf z}$\in${\Bbb O}
\beq                                                
d{\bf F}={\bf F}({\bf z}+d{\bf z})-{\bf F}({\bf z})
\eeq
be an infinitesimal increment (differential) of {\bf F}({\bf z}),
corresponding to a differential of a {\Bbb B}\,-variable $d{\bf z}$
and according to the usual
Euclidean metric $\rho^2=\sum_\mu|z^\mu|^2$. Then we come to the
following definition.

The function (1) {\bf F}({\bf z}) is said to  be {\Bbb B}\,-{\it
differentiable\/} in  some
domain {\Bbb O}$\subset${\Bbb M} if for $\forall${\bf z}$\in${\Bbb O} there
are some {\bf G}({\bf z}), {\bf H}({\bf z}) such that
{\bf the differential (3) may be presented in the invariant form}
\beq                                     
d{\bf F}={\bf G}({\bf z})\ast d{\bf z}\ast{\bf H}({\bf z}),
\eeq
i.e.\ only through the operation of multiplication in {\Bbb B}\,.

For the commutative algebra of complex numbers, from (4) the Cauchy-Riemann
(CR) equations follow in the coordinate  representation, ${\bf F}^\prime=
{\bf G}\ast{\bf H}$ being  a derivative  of {\bf F}({\bf z}). So the
relation (4)
naturally generalizes the CR equations to the case of a non-commutative
associative {\Bbb B}\,-algebra. Eqs.\,(4) will be further designated as GCRE
(in the invariant form).

A detailed study of {\Bbb B}\,-differentiability  and  analyticity,
based on GCRE (4), may  be  found  in  [3], and a review  of  other
approaches in [9]. The most  profound is perhaps Fueter's
work [10]; G\"ursey et al.\ [11] applied it  within  the   $d=4$  gauge  
and chiral theories (see also [12]).

\section{Spinor splitting and the eikonal eq\-uation}   

Let us turn now to the construction of field  theory,  based  on
the concept of {\Bbb B}\,-differentiability. Consider a subspace $\mbox{{\Bbb
M}}_+\subset\mbox{\Bbb M}$ of the
points with real coordinates
${\bf x}=\{x^\mu\}=\{z^\mu:{\rm Im}\,(z^\mu)=0\}$, or  else  the
subspace  of  Hermitian  matrices  with  elements ${\bf z}^+={\bf z}$.
The {\Bbb B}\,-norm
$N^2({\bf z})={\rm Det}\,({\bf z})$ then generates on $\mbox{{\Bbb M}}_+$ the
real Minkowski metric
\bearr                                       
N^2({\bf x})={\rm Det}\,({\bf x})=uv-pw \nnn
\cm = \left(x^0\right)^2-\left(x^1\right)^2
     -\left(x^2\right)^2-\left(x^3\right)^2
\ear
so $\mbox{\Bbb M}_+$ may be identified with the physical ST. Solutions
to (4) on {\Bbb M} may be obtained by analytic continuation from
$\mbox{\Bbb M}_+$. We will return to a detailed study of the relation
between {\Bbb M} and $\mbox{\Bbb M}_+$ in Sec. 7.

It is now evident that  the {\Bbb B}\,-differentiable  functions {\bf
F}($x$), realizing the mappings ${\bf F}\colon\mbox{\Bbb M}_+\to\mbox{\Bbb
M}$, should be considered  as a fundamental physical field; its spinor
nature  will  be  seen  below.  We will  assume {\bf the dynamics of a basic
F-field  to be  completely determined by the GCRE (4) with} ${\bf z}={\bf
x}\in\mbox{\Bbb M}_+$, i.e.\
\beq                                      
d{\bf F}={\bf G}(x)*d{\bf x}*{\bf H}(x)
\eeq
Except direct  physical identifications  of the abstract  variables, in what
follows no other  assumptions  will  be  necessary.

Let us rewrite now the matrices {\bf F}, {\bf H} in (6) in the form
\beq                                     
{\bf F}=\|\psi(x),\eta(x)\|,\quad{\bf H}=\|\alpha(x),\gamma(x)\|
\eeq
each of $\psi,\eta$ and $\alpha,\gamma$ being a matrix-column with two
components; the columns transform independently through left  multiplication.
Then (6) splits into a pair of equations:
\beq                                     
d\psi={\bf G}*d{\bf x}*\alpha,\quad d\eta={\bf G}*d{\bf x}*\gamma.
\eeq

From (7) and (8) it follows that  each solution  to  (6) may  be
presented in the  form ${\bf F}(x)=\|\psi^\prime(x),\psi^{\prime\prime}(x)\|$
where $\psi^\prime,\psi^{\prime\prime}$ are two  arbitrary
solutions of the unique irreducible equation
\beq                                     
d\psi={\bf G}(x)*d{\bf x}*\alpha(x).
\eeq

The functions $\psi(x),\alpha(x)$ belong  to  the {\it left-side  ideal\/} of
the Clifford algebra $\mbox{\Bbb B}=\mbox{C(3,\,0)}$ and are therefore
obviously {\it 2-spinors}. A conjugated spinor reduction of (6)  is  also
possible,  if  the  row splitting of {\bf G}($x$) is used; a double reduction
may be realized as well.

These properties stand side by side with  the  widest  symmetry
group of Eqs.\,(4) or (6), including the transformations
\beq                                                          
\left.\begin{array}{@{}r@{\,}c@{\,}lr@{\,}c@{\,}l}
{\bf z}&\to&{\bf m}*{\bf z}*{\bf n}^{-1},&{\bf F}&\to&{\bf k}*{\bf
F}*{\bf l}                                        \\[2pt]
{\bf G}&\to&{\bf k}*{\bf G}*{\bf m}^{-1},&{\bf H}&
\to&{\bf n}*{\bf H}*{\bf l}
\end{array}\right\},
\eeq
{\bf m},\,{\bf n},\,{\bf k},\,{\bf l} being arbitrary
constant biquaternions of
unit norm (neglecting the dilatations ${\bf z}\to\lambda{\bf z}$,
$\lambda\in\mbox{\Bbb C}$), ${\bf m}^{-1},{\bf n}^{-1}$ are the inverse ones.

{\bf Z}-transformations  in (10) define  a  6{\Bbb C}\,-parameter  group  of
rotations SO\,(4,\,{\Bbb C});
the restriction of this group to $\mbox{\Bbb M}_+$
(with ${\bf n}^{-1}={\bf m}^+$)
leads to the Lorentz transformations for {\bf x}. Now, if we put in (10)
${\bf k}={\bf n}$, ${\bf l}={\bf m}^{-1}$, the functions {\bf F}, {\bf G},
{\bf H}  manifest  their  nature  as
4-vectors (${\bf F}\to{\bf n}*{\bf F}*{\bf m}^{-1}$ etc.). However, when
${\bf k}={\bf n}$, ${\bf l}={\rm Ident.}$, {\bf G}
transforms as a 4-vector, for {\bf F}  and {\bf H} we have ${\bf F}\to
{\bf n}*{\bf F}$, ${\bf H}\to{\bf n}*{\bf H}$, preserving the structure of
the spinor splitting (8).  Moreover, a double
row-column splitting of (6) corresponds to the case ${\bf k}={\bf l}={\rm
Ident.}$, when
{\bf F}($x$) has to be considered as a scalar, while
{\bf G} and {\bf H} transform
as conjugated spinors ${\bf G}\to{\bf G}*{\bf m}^{-1}$, ${\bf H}\to{\bf n}*
{\bf H}$.

Let us return now to the dynamical consequences of GCRE (6) and (9).
Using the {\it Fiertz identity\/} and the double row-column splitting of (8),
for every matrix component $\psi={\bf F}_{AB}$; $A,B\!=\!1,2$ of {\bf F}-field
we get~[13]
\beq                                                             
(\partial_0\psi)^2-(\partial_1\psi)^2-(\partial_2\psi)^2-(\partial_3\psi)^2=0.
\eeq

Hence, {\bf every  matrix  component  of
a {\Bbb B}\,-differ\-en\-tiable function
satisfies the nonlinear,  Loren\-tz  invariant,  complexified  eikonal
equation} (11). For the {\Bbb B}\,-algebra it plays a role similar to that
of the Laplace equation  in  complex  analysis;  as  for  physics,  its
fundamental properties (for $\psi\in\mbox{\Bbb R}$) were emphasized by
V.A. Fock~[14].

\section{{\Bb B}\,-electrodynamics.~Self-duality co\-nditions and Maxwell
equations}                     

We shall further restrict ourselves  to  the  case  of the spinor
equality $\alpha(x)=\psi(x)$ in (9), i.e.\ to the fundamental equation
\beq                                                      
d\psi={\bf G}(x)*d{\bf x}*\psi(x).
\eeq
For (12) the global continuous symmetries (10)  are  reduced  to
the transformations of the Lorentz group
\beq                                                       
{\bf x}\to{\bf m}*{\bf x}*{\bf m}^+,\ \psi\to{\bf s}\psi,\
\overline{\bf G}\to{\bf m}*\overline{\bf G}*{\bf m}^+,
\eeq
where ${\bf s}=({\bf m}^+)^{-1}$, $\overline{\bf G}$ is a {\Bbb
B}\,-conjugated field: $\overline{\bf G}*{\bf G}=({\rm Det}\,{\bf G})^2$.

So relativistic invariance is ensured, and the conjugated  field
$\overline{\bf G}(x)$  forms a 4-vector. Later on $\overline{\bf G}(x)$ will
be regarded as a {\Bbb C}\,-valued
matrix of electromagnetic (EM-) 4-potential {\bf A}($x$). Precisely, we set
\beq                                                          
A_\mu(x)=2\overline{G}_\mu(x)\equiv 2G^\mu(x).
\eeq
Such an identification will be justified further by  its  dynamic  and
geometric consequences, as well as by the  establishment  of  gauge
invariance of (12). Therefore, the latter {\bf can be considered  as the
basic equations of {\Bbb B}\,-electrodynamics}, i.e., some type of classical
{\it spinor electrodynamics}, generated by solely the GCRE-structure.

Written in components, Eqs. (12) form the  set of differential equations
\beq                                             
\left.\begin{array}{@{}l@{\ }l@{\ }l@{\ }l@{}}
\partial_uf\!=\!G^u\!f,&\partial_pf\!=\!G^w\!f,&\partial_wf\!=\!G^u\!h,&
\partial_vf\!=\!G^w\!h\\[2pt]
\partial_uh\!=\!G^p\!f,&\partial_ph\!=\!G^v\!f,&\partial_wh\!=\!G^p\!h,&
\partial_vh\!=\!G^v\!h
\end{array}\right\}.
\eeq
Here $f(x)$ and $h(x)$ are the components of a 2-spinor field $\psi(x)$, and
$\partial$
denotes a partial derivative with respect to the corresponding DeWitt
coordinate.

The equations for the EM field follow  from  the over-det\-er\-mined system
(15) as its {\it integrability (compatibility) conditions}
\[
\partial_\mu(\partial_\nu\psi)-\partial_\nu(\partial_\mu\psi)=0,\quad
\psi=\{f(x),h(x)\}.
\]
Assuming then both $f(x),\ h(x)\not\equiv 0$ (otherwise
we would have obtained the same final results), we obtain after derivation
\beq                                                      
\left.\begin{array}{@{}r@{\,}c@{\,}l@{\ }r@{\,}c@{\,}l@{}}
\partial_uA_w-\partial_pA_u&=&0,&\partial_wA_w-\partial_vA_u&=&
\frac{1}{2}\,{\rm Det}\,{\bf A}\\[2pt]
\partial_vA_p-\partial_wA_v&=&0,&\partial_pA_p-\partial_uA_v&=&
\frac{1}{2}\,{\rm Det}\,{\bf A}
\end{array}\right\},
\eeq
$A_\mu(x)$ being the EM potentials (14)
and ${\rm Det}\,{\bf A}=A_uA_v-A_pA_w$. Going back to the
Cartesian coordinates, we observe that Eqs. (16)
are equivalent to the {\it self-duality conditions\/} (SDC)
\beq                                    
\vec{\cal P}\equiv\vec{\cal E}+i\vec{\cal B}=0
\eeq
for the {\Bbb C}\,-valued electric $\vec{\cal E}=\{{\cal E}_a\}$ and magnetic
$\vec{\cal B}=\{{\cal B}_a\}$ components of the EM field strength tensor
\beq                                               
{\cal F}_{\mu\nu}=\partial_\mu A_\nu-\partial_\nu A_\mu;
\eeq
here
\beq                                                  
{\cal E}_a={\cal F}_{{\scriptscriptstyle 0}a},\ {\cal B}_a=
{\textstyle\frac{1}{2}}\,\varepsilon_{abc}
{\cal F}_{bc};\ a,b,c,\dots=1,2,3.
\eeq
In addition to (17), from (16) we have
\beq                                                   
{\cal D}\equiv\partial_\mu A^\mu + 2A_\mu A^\mu=0,
\eeq
i.e.\ an {\it inhomogeneous Lorentz condition}.

Combined with the definitions (18) and (19), the {\bf SDC (17)  lead  then
to the Maxwell equations in free space}
\beq
\partial_\nu{\cal F}^{\mu\nu}=\partial_\nu\left({\textstyle\frac{1}{2i}}
\varepsilon^{\mu\nu\rho\lambda}{\cal F}_{\rho\lambda}\right)=0.
\eeq
So the Maxwell equations represent nothing but the {\it consistency
conditions\/} of a basic GCRE-system and {\bf are satisfied identically 
for each  solution to the latter}. The inverse statement generally does 
not take place!

Now, it is easy to see that, according to the SDC (17),  the {\it energy-%
momentum density\/} of a complex-valued EM-field turns to zero. Therefore,
we ought to define the {\bf physical} fields $\vec E$, $\vec B$ through the
real (Re) or imaginary (Im) parts of (19). For geometric reasons (see part 6),
we prefer
\beq                                                             
\vec E=2{\rm Re}\,(\vec{\cal E}\,),\quad\vec B=2{\rm Re}\,(\vec{\cal B}\,).
\eeq
The {\Bbb R}-valued vectors $\vec E$, $\vec B$ satisfy the linear Maxwell
equations as well.
However,  they  are  mutually  independent  (contrary  to  (19))  and
create a non-zero energy-momentum density ($W$, $\vec P$) of the usual form
\beq                                                             
W\sim\left(|\vec E|^2+|\vec B|^2\right),\quad\vec P\sim[\vec E\times\vec
B].
\eeq
Moreover, an infinite series of conservation laws can be obtained  for
(15) using routine procedures (see [15] for an example).

\section{Coulomb field as a stereographic map. Electric charge quantization}

Let us  search  now  for solutions  to the {\Bbb B}\,-electrodyna\-mic
equations (15). Each of the two components $f(x)$, $h(x)$ of the spinor field
in (15) satisfies the {\Bbb C}\,-eikonal equation (11). Starting from one
of its solutions, all the other quantities, including the EM potentials (14),
should be derived. In particular, the  wave-like  solutions
of (11) lead to EM fields, identical to the usual EM  waves [3,\,13].

Notice now that {\bf the eikonal equation (11) possesses  a  wonderful
invariance property} under the transformations
\beq                            
\psi(x)\to\Phi\left(\psi(x)\right)
\eeq
with an arbitrary ({\Bbb C}\,-differentiable) function $\Phi(f)$.
Accordingly,
one can easily verify the {\it gauge invariance\/} of the basic system (12)
(and, therefore, (15)) of a special type:
\beq
\psi(x)\to\psi(x)\alpha(\psi),\ A_\mu(x)\to A_\mu(x)\!+\!\partial_\mu
\ln\alpha(\psi),
\eeq
$\alpha(\psi)$ being an arbitrary scalar function of the $\psi$ components
$f(x)$, $h(x)$.

The {\Bbb C}\,-structure of the eikonal equation (11)  essentially  enlarges
the spectrum of its solutions. The most important are certainly  two
static solutions found in [3]:
\beq                           
\left.\begin{array}{@{}l}
f^+={\displaystyle\frac{x^1+ix^2}{r+x^3}}=\tan\left(\frac{\theta}{2}\right)
\exp\,(i\varphi)\\[7pt]
f^-={\displaystyle\frac{x^1-ix^2}{r-x^3}}=\cot\left(\frac{\theta}{2}\right)
\exp\,(-i\varphi)
\end{array}\right\},
\eeq
where $\{r,\theta,\varphi\}$ are usual spherical
coordinates on $\mbox{\Bbb R}^3$. From a geometric
point of view,  the  expression  (26) {\bf corresponds  to the stereographic
projection} $\mbox{\Bbb S}^2\to\mbox{\Bbb C}$ of a unit 2-sphere onto the
{\Bbb C}\,-plane (from the south
(+) or north (-) poles, respectively). Substituting  (26)  into  (15),
(14), we get after trivial integration:
\beq                              
\left.\begin{array}{@{}r@{\,}c@{\,}l@{}r@{\,}c@{\,}l@{}}
f(x)&=&f^\pm(\theta,\varphi),&h(x)&=
&\left[f^\pm(\theta,\varphi)\right]^2\\[5pt]
A_u&=&\mp{\displaystyle\frac{1}{r}},&A_v&=&\pm{\displaystyle\frac{2}{r}}\\[5pt]
A_p&=&{\displaystyle\frac{e^{-i\varphi}\!\left(\tan\frac{\theta}{2}
\right)^{\mp 1}}{r}},&
A_w&=&-{\displaystyle\frac{2e^{i\varphi}\!\left(\tan\frac{\theta}{2}
\right)^{\pm 1}}{r}}
\end{array}\right\}
\eeq
or, for spherical components of the {\Bbb C}\,-valued 4-potential
\bearr                  
A_0=\pm\frac{1}{2r},\quad A_r=-\frac{1}{2r},\quad A_\theta=\mp
										iA_\varphi=\nnn
 -\frac{1}{2r}\cot\theta\pm\frac{3}{2r\sin\theta}.
\ear

Now, a transition to the physical vectors of EM-field
strengths (22) shows that the {\it magnetic monopole\/} and gradient-like
terms in (28) disappear and we get
\beq
E_\theta=E_\varphi=B_r=B_\theta=B_\varphi=0,\quad E_r=\pm\frac{1}{r},
\eeq
i.e.\ {\bf the Coulomb law with a fixed value of electric charge} $q=\pm 1$.

Whereas the stereographic projection (26) and the transformations  (24)
realize the conformal mappings $\mbox{\Bbb S}^2\to\mbox{\Bbb C}$, $\mbox{\Bbb
C}\to\mbox{\Bbb C}$ respectively, EM fields
behave by (25) in a gauge invariant manner,  and  the
electric charge remains quantized. An exceptional role of conformal
mappings in algebrodynamics has been clarified in [3] (chapter 1).

$q$-quantization is a crucial  point  for {\Bbb B}\,-electrody\-na\-mics;
a fundamental significance of this problem has been evident to  Dirac,
Eddington, Wheeler and other grands. In orthodox
field theory the $q$-quantization is postulated rather than explained.
The most elegant approach to this problem is produced, perhaps, by
multidimensional ST theories [16]; {\Bbb B}\,-electrodynamics presents
another possibility.

In our approach the algebraic and purely classical origin of
$q$-quantization becomes evident. The fact is that the initial GCRE  are
not invariant under the scaling ${\bf A}\to\lambda{\bf A}$, contrary to
the linear Maxwell
equations. We suppose, however,  that  the  phenomenon  of {\bf ``algebraic
$q$-quantization''} should have deeper topological  reasons;  we
hope to discuss them in the future [20].

\section{Spinor connection and Weyl-Cartan geometry of ST}   

The fundamental equation of {\Bbb B}\,-electrodynamics (12) may be presented
in the form
\beq                              
\partial_\nu\psi={\bf\Gamma}_\nu(x)\,\psi(x),
\eeq
with
\beq                                 
{\bf\Gamma}_\nu(x)={\bf G}(x)*\sigma_\nu
\eeq
being a {\it 2-spinor  connection\/}  of special type. The initial GCRE,
corresponding to (30), have the matrix form (6) with ${\bf H}(x)={\bf F}(x)$,
i.e.
\beq                                    
d{\bf F}={\bf G}(x)*d{\bf x}*{\bf F}(x),
\eeq
or, in a 4-vector representation [3]
\bearr                                     
\partial_\nu F^\mu=\Gamma^\mu_{\nu\!\rho}(x)\,F^\rho, \\  \lal
\Gamma^\mu_{\nu\!\rho}(x)=2(A_\nu\delta^\mu_\rho+A_\rho\delta^\mu_\nu-
A^\mu\eta_{\nu\!\rho}-i\varepsilon^\mu_{\cdot\nu\!\rho\alpha}A^\alpha)
\ear
where $\delta^\mu_\nu$, $\eta_{\mu\nu}$, $\varepsilon_{\mu\nu\!\rho\lambda}$
are the Kronecker, Minkowski  and  Levi-Civita
tensors, respectively, and $A_\mu(x)$ are {\Bbb C}\,-valu\-ed potentials
(14).

Thus in the basic electrodynamic case the initial GCRE system (6) is
equivalent to the defining equations  (32)  of the {\it covariantly  constant
vector fields\/} $\{F^\mu(x)\}$ on a {\Bbb B}\,-manifold  with a
``dynamically created'' effective {\bf geometry  of  Weyl-Cartan type},
represented by the affine connection (34). Note that the {\Bbb C}\,-vector
$A_\mu(x)$ completely determines both the Weyl part of (34) and its
torsion  structure.  A generalization by introduction of a Riemann
metric structure is natural as well.

To obtain the ST geometry induced by (34), let us pass
from ${\bf F}(x)$ to the unitary field ${\bf U}(x)={\bf F}*{\bf F}^+$.
Using (32), we get
\beq                                             
\partial_\nu U^\mu=\Delta^\mu_{\nu\!\rho}(x)\,U^\rho(x),
\eeq
with the {\Bbb R}-valued connection
\beq                                                
\Delta^\mu_{\nu\!\rho}(x)=2(a_\nu\delta^\mu_\rho+a_\rho\delta^\mu_\nu-
a^\mu\eta_{\nu\!\rho}-\varepsilon^\mu_{\cdot\nu\!\rho\alpha}b^\alpha),
\eeq
where $a_\mu(x)$ and $b_\mu(x)$
are real and imaginary parts of the potentials $A_\mu(x)$.

A connection similar to (36) has  been  introduced  in Ref.\,[17]  from
physical considerations; in [18] it was shown  to  be {\bf the  only ST
connection compatible with a spinor bundle structure} with the conventional
notion of a covariant spinor derivative. In our approach  these  results
follow from the GCRE structure alone.

However,  the {\it torsion  field\/} $b_\mu(x)$ in (36) satisfies the Maxwell
equations, as well as the {\it non-metricity field\/} $a_\mu(x)$. By the
key
Ansatz (28), precisely the Weyl part $a_\mu(x)$ corresponds to the ordinary
Coloumb electric field, justifying the previous identification of the EM
field with the real part of the {\Bbb C}\,-field.

As for the imaginary part $b_\mu(x)$, for (28) it has the magnetic monopole
form
\beq                          
b_0=b_r=b_\theta=0,\quad b_\varphi=\mp\frac{1}{2r}\cot\theta-
\frac{3}{2r\sin\theta};
\eeq
we thus come  to  an  exotic {\bf geometric interpretation  of  magnetic
monopoles as a ST torsion} (with a totally  antisymmetric  tensor
structure). Accordingly, the field $b_\mu(x)$ cannot  appear in the
equations of geodesics. If we assume the latter to present the laws
of test particle motion, then {\bf monopoles should} have no
effect on it and therefore {\bf be entirely unobservable}!

Let us now return to the study of the primary {\Bbb C}\,-geometry of the
{\Bbb B}\,-space.
The integrability conditions for the irreducible spinor equation (30) may
be written in the form
\beq                                 
{\bf R}_{\mu\nu}\psi(x)=0,
\eeq
with
\beq                                    
{\bf R}_{\mu\nu}=\partial_{[\mu}{\bf\Gamma}_{\nu]}-[{\bf\Gamma}_\mu,
{\bf\Gamma}_\nu]
\eeq
being the {\it curvature tensor\/} in the matrix representation. For
its self-dual components
\beq                                      
(\vec{\bf R})_a={\bf R}_{{\scriptscriptstyle 0}a}+{\textstyle\frac{i}{2}}
\varepsilon_{abc}{\bf R}_{bc}
\eeq
with the connection of the form (31), we get
\beq                                            
\vec{\bf R}=\vec{\cal P}+{\cal D}\vec\sigma-i[\vec{\cal P}\times\vec\sigma\,].
\eeq
Here the quantities
\[
\vec{\cal P}=\vec{\cal E}+i\vec{\cal B},\quad{\cal D}=\partial_\mu A^\mu
+2A_\mu A^\mu
\]
coincide with (17) and (20), respectively and therefore vanish
along with the entire self-dual tensor (40).

It is easy to see that $\cal D$ is  proportional  to  the  curvature
invariant ${\cal D}=6\eta^{\mu\nu}R^\alpha_{\mu\alpha\nu}\ (=0)$.
So the {\bf{\Bbb B}\,-space  appears  to  be  an  self-dual  space  with
zero scalar curvature}. Now, if there are two {\bf linearly  independent}
spinor solutions of (30), then, as follows from (38),
\beq                                                 
{\bf R}_{\mu\nu}(x)=0,
\eeq
i.e.\ a trivial case of flat geometry and zero field strengths.

To avoid that, the primary {\Bbb B}\,-field ${\bf F}(x)$ in (32) should
split into
two  spinors $\psi^\prime(x)$, $\psi^{\prime\prime}(x)$ (see (9)),
proportional to  each  other; therefore, we have
\beq                          
{\rm Det}\,{\bf F}(x)=0,
\eeq
and the field ${\bf F}(x)$ takes the values on the subspace of null divisors
of the {\Bbb B}\,-algebra, or, physically, {\bf on the complex ``light
cone''}.

The null {\Bbb B}\,-fields are the most fundamental  objects  throughout
the AD approach as a whole. However, they can exist only on manifolds
with an indefinite metric signature. So the {\bf pseudo-Euclidean structure}
of the World should not be postulated within the AD app\-roach, but {\bf is
just a necessary condition of nontrivial dynamics (and effective
geometry)}.

\section{Yang-Mills fields and the {\Bb C}\,-structure of space-time}     

Now we will demonstrate  that the  Yang-Mills  (YM)  gauge
fields also appear in theory in rather a natural way. To see that, let
us separate the {\it trace-free part\/} in the basic spinor connection (31)
\beq                                    
{\bf\Gamma}_\nu(x)={\bf G}(x)*\sigma_\nu={\textstyle\frac{1}{2}}
\bigl(A_\nu(x)+ {\bf N}_\nu(x)\bigr).
\eeq
Then the zero component $A_\mu(x)$
coincides  with the {\Bbb C}\,-potentials (14) of the EM field, and the
trace-free part ${\bf N}_\mu(x)$ can be expressed in its terms in a linear
way:
\bearr                        
{\bf N}_\mu(x)=N_\mu^a(x)\,\sigma_a;
\quad N^a_{\scriptscriptstyle 0}=A_a(x),\nnn
\quad N^a_b=\delta_{ab}
A_{\scriptscriptstyle 0}(x)-i\varepsilon_{abc}A_c(x).
\ear
The quantities ${\bf N}_\mu(x)$ can be regarded as the matrix potentials of
some {\Bbb C}\,-valued gauge field; its strength corresponds to the traceless
part of the curvature tensor (39) and may be written as usual:
\beq                                                          
{\bf L}_{\mu\nu}={\cal L}^a_{\mu\nu}(x)\,\sigma_a=\partial_{[\mu}
{\bf N}_{\nu]}-[{\bf N}_\mu,{\bf N}_\nu].
\eeq

We see now that the self-dual part of (46) coincides with the traceless
part of the tensor (41) and, in view of (17) and (20), we have again
\beq                                             
{\bf L}_{\mu\nu}+{\textstyle\frac{i}{2}}\varepsilon_{\mu\nu\!\rho\lambda}
{\bf L}^{\rho\lambda}=0.
\eeq
From (47) and the {\it Bianchi identity\/}, the YM equations
follow immediately in a usual way:
\beq                                                            
\partial_\nu{\bf L}^{\mu\nu}=[{\bf N}_\nu,{\bf L}^{\mu\nu}].
\eeq
So we can indeed consider the field ${\bf N}_\nu(x)$ as a {\Bbb C}\,-valued
YM field
of a special structure (45). From (38) for any non-trivial $\psi(x)$ we
obtain, in addition,
\beq                            
{\rm Det}\,{\bf R}_{\mu\nu}=0;
\eeq
written in components, this condition leads to an expression of the EM field
strength in terms of the YM ones (for each $[\mu\nu]$ separately):
\beq                                                     
{\cal L}^a_{\mu\nu}{\cal L}^a_{\mu\nu}=({\cal F}_{\mu\nu})^2.
\eeq
So {\bf the EM field may be regarded as a modulus of the YM triplet  field
in the isotopic complexified 3-space.}

Contrary to EM fields, the YM ones cannot be split  into  real  and
imaginary parts (due to the nonlinearity  of the  YM equations)  and
therefore are essentially {\Bbb C}\,-valued. This seems quite natural in
connection with the
pseudo-Euclidean structure of the ST (the duality operator is known to
have imaginary eigenvalues in the Lorentz signature). Since the
self-duality conditions play a crucial role  both in orthodox  field
theory and in AD, we come again to the {\bf conjecture on a {\Bbb
C}\,-analytic structure of real physical ST}. This possibility, discussed
repeatedly
within the frames of GRT, the twistor and string programs, as well as within
the {\it binary geometrophysics\/} approach [19], seems to be inevitable in
AD in view of the non-existence of a {\Bbb R}-valued STA with an Aut group
isomorphic to the Lorentz one. Thus, we  suppose that close  connections
between the field equations nonlinearity and the {\Bbb C}\, structure of
ST do exist, as well as its noncommutative quaternion structure (see
Sections 1 and 3 for the latter).

Moreover, {\bf we may think of the {\Bbb C} structure as some natural
way of ST dimension enlarging} (namely, doubling), just  in  the sense  of
Kaluza-Klein theories. As  for  physics,  such  an  effect should  be
essential at high energies; asymptotically, in the linear  approximation,
the ST {\Bbb C}\, structure should split into the Minkowski space observed
plus a conjugated one. The same is done by the field  {\Bbb C}\,-structure:
it exhibits  a reduction to a linear {\Bbb R}-valued EM-field (doubled
through the SDC (17), too).

Generally,  we assume  the  existence  of a biquaternion  (i.e.\
complex-quaternion) algebraic structure of  ST  and  field  manifolds
consistent with each other. The {\bf non-commutativity} of such a
{\Bbb B}\,-algebra
{\bf results in the nonlinearity of fundamental dynamics}: the GCRE (6),
{\bf
as well as its P- and even T-noninvariance}  (the  connections  similar
to (35), (36) are efficiently employed by V.G. Krechet for a 5-geometrical
description of electroweak interaction). Nevertheless, here  the
usual reversible dynamics of gauge fields has been obtained in Sections
4 and 7; this latter should be regarded as nothing more but some  ``trace''
of a primary {\Bbb B}\,-structure, responsible for
interactions, the ``time arrow'' and the left-right preference
on  the  Minkowski  ST. We  expect  an
extensive presentation of our views of these problems  as well
as numerous generalizations of the AD approach.

In  conlusion,  peculiar  correlations  between  AD  and  binary
geometrophysics (BG) [19] should be noted.  Both  of  the  approaches
start from some abstract exceptional  algebraic  structures  and  deal
with either basic relations (in BG), or special mappings  (in
AD). In both theories {\bf zero determinant structures} (see (43), (49))
{\bf are
of particular importance}. Finally, the ideas of {\it multipoint geometries\/}
[19,\,20] originate from purely algebraic considerations and should  find
their place  in  AD  as  well.  It  seems  plausible  that  other
deep interrelations will be found out in future.

We see that the simplest AD model, based on the conditions of {\Bbb
B}\,-differentiality alone, naturally contains the geometric, spinor-gauge
and discrete structures, capable of solving the charge quantization and
monopole problems. Within this model, the Coulomb law gains an exotic
geometrical meaning, and the {\Bbb C}\,-eikonal equation becomes a
fundamental equation of field dynamics. Related problems (in particular,
the problem of motion law and many-sources distributions) are yet to be
solved.

\Acknow{I am grateful to D.V.~Alexeevsky, B.V.~Medvedev and especially  to
Yu.S.~Vladimirov for helpful  advice  and (Yu.S. Vladimirov) for
organizational support.}

\end{document}